\documentclass[10pt,pra,aps,showpacs,twocolumn,unsortedaddress]{revtex4-1}

\usepackage{graphicx}
\usepackage{epstopdf}
\usepackage{hyperref}

\begin{document}

\title{Lattice gas automaton modelling of a vortex flow
meter: Strouhal-Reynolds number dependence}

\author{Vaidas Juknevi\v{c}ius}
\email{vaidas.juknevicius@tfai.vu.lt}
\author{Jogundas Armaitis}
\affiliation{Institute of Theoretical Physics and Astronomy,
Vilnius University,
Saul\.{e}tekio al. 3, LT-10222 Vilnius, Lithuania}

\begin{abstract}
Motivated by recent experimental and computational results 
concerning the three-dimensional
structure of vortices behind a 
vortex shedding flow meter [M. Reik et al., Forsch. Ingenieurwes.
74, 77 (2010)], 
we study the Strouhal-Reynolds number dependence in the 
vortex street in two
dimensions behind a trapezoid-shaped object 
by employing two types of Frisch-Hasslacher-Pomeau (FHP) models. 
Our geometry is intended to reproduce the operation of 
the vortex shedding flow meter 
in a two-dimensional setting, thus preventing the formation 
of three-dimensional vortex structure. 
In particular, we check if the anomalous Reynolds-Strouhal number dependence
reported for three dimensions can also be found in our two-dimensional 
simulation.
As we find that the Strouhal number is nearly independent of the Reynolds 
number in this particular setup, 
our results provide support for the hypothesis that 
three-dimensional flow structures 
are responsible for that dependence, thus hinting at the importance of 
the pipe diameter to the accurate operation of industrial vortex flow meters.
\end{abstract}

\maketitle

\section{Introduction}
\label{section:intro}
Hydrodynamic theories have been studied for a long time, but they
still provide new insights in various problems of physics and 
engineering, from nondissipative currents in ultracold atomic vapour \cite{book:pitaevskii} to stability of tall buildings \cite{article:irwin_2010},
in addition to posing some extremely challenging questions along the way \cite{link:claymath}. 
Several branches of the field remain
particularly vigorous, including studies of flow instabilities \cite{book:sengupta}. Besides
being of fundamental importance, flow instabilities in general, and the renowned
K\'arm\'an vortex street 
\cite{article:mallock,article:benard}
in particular can be readily observed in everyday life \cite{book:karman}, and also has important practical uses.

The vortex shedding flow meter stands out as a direct industrial application of the phenomenon of
the K\'arm\'an vortex street. This device consists of a blunt object positioned inside a pipe,
and a detector of vortices. As liquid (or gas)
flows through the pipe sufficiently fast, a vortex street forms downstream from the blunt object. Since
the vortex shedding frequency is dependent on the hydrodynamic properties of the flow,
the signal of the vortex detector can be converted to the velocity measurement of the flow. This type
of a device is uniquely suitable for operation in an industrial setting, as it is fully contained
inside the pipe, has no moving parts, and is both robust and reliable \cite{book:boyes}. 

It turns out that even in this seemingly mundane setting of a tested industrial application,
novel physics can be uncovered. In particular, it has been recently suggested that the vortex
pattern that forms downstream from the vortex flow meter has a three-dimensional structure
\cite{article:reik_et_al_2010}, in the shape of so-called horseshoe vortices. In turn,
this spatial structure alters the flow in the pipe, introducing an anomalous
relation between the Reynolds and the Strouhal numbers. This anomalous relation
violates the main operating assumption of the flow meter, therefore leading to
inaccurate flow measurements in certain regimes. In this paper, we provide an
additional check if the spatial structure is indeed to blame for the anomalous
relation by numerically studying the same problem in a two-dimensional geometry,
where the horseshoe vortices cannot form.

The traditional approach to hydrodynamics, namely solving the Navier-Stokes equations, 
is considered to be both analytically and computationally complicated. Analytical solutions 
of hydrodynamic problems are possible only in limited number of cases of flows at small Reynolds numbers and in relatively simple geometries.
 Some hydrodynamic instabilities can be investigated analytically and numerically 
using dynamical systems approaches. 
{
For example, 
expansion around instabilities may result in equations that are simpler than Navier-Stokes equations, but which are nevertheless able to reproduce the formation of hydrodynamic patterns \cite{book:ds_appr_turb_1998,{book:hydro_instab_trans_turb_1981}}. Also, in some cases, weakly or moderately turbulent flows can be described using the so-called shell models that consist in replacing the partial differential equation with a system of coupled ordinary differential equations via discretization and truncation in Fourier space   \cite{book:ds_appr_turb_1998}.  
}
However, for extremely high Reynolds numbers, because of the large number of relevant degrees of freedom due to the wide range of scales, the traditional descriptions of fully developed turbulence employ statistical methods \cite{book:monin1,book:monin2} and phenomenological models \cite{{article:kolmogorov41},{article:kolmogorov62}}.

An altogether different approach to hydrodynamic problems are the so-called lattice gas models \cite{review:wolfram,{book:lattice-gas_hydro_2001},thesis:wylie_1990}.
These models belong to a wide class of discrete systems known as cellular
automata. They have a direct physical interpretation. Namely, point particles
occupy nodes of a lattice with the possibility to jump from one node to one
of its neighbouring nodes in a single time step. In most of these models
the particles move with a single speed in one of several directions.
Collisions of the particles occur at the nodes, and are executed
according to some simple logical rules.
Remarkably, if the lattice has proper symmetries and the
collision rules satisfy relevant conservation laws (e.g., momentum and
energy conservation), then the global behaviour
of the system in a coarse-grained picture will closely resemble
the flow of a fluid~\cite{book:lattice-gas_hydro_2001,article:fhp}. 

In this work, two different rule sets of the seven-particle Frisch-Hasslacher-Pomeau (FHP, \cite{article:fhp})
model have been used in order to simulate the vortex flow meter \cite{article:reik_et_al_2010}.
{
These FHP models have been successfully employed to attack diverse
problems, including nucleation in supersaturated liquids \cite{article:hickey}, sand dune growth \cite{article:gao}, as well as
flows on curved surfaces with dynamical geometry 
\cite{article:klales}, in addition to shedding insight on 
various aspects of hydrodynamics \cite{book:lattice-gas_hydro_2001}. Hence, 
even though more elaborate methods to address fluid dynamics are present 
(for example, the lattice Boltzmann equation \cite{book:succi}), in our case we have 
employed the FHP
lattice gas model in order to minimize the computation effort, while still 
obtaining reliable results.
}

{ The structure of the paper is as follows.}
Section \ref{section:models} introduces the FHP lattice gas models used for the simulations, together with the definitions of Reynolds and 
Strouhal numbers in this framework. 
Then, in Section \ref{section:simulation}, the main results are presented and discussed. 
Finally, Section \ref{section:conclusions} summarizes the results and draws some conclusions, in addition to discussing several
promising directions for future investigations.
\section{Methods of simulation}
\label{section:models}

In this section we briefly describe the lattice gas automata in general, and the FHP models in particular. Even though several excellent
resources on these subjects are available (see, e.g., Refs.\ \cite{book:lattice-gas_hydro_2001} and \cite{thesis:wylie_1990} as well as the references within them), we summarize the most important
aspects of the methods employed in order to make our discussion self-contained.

The lattice gas automata consist of discrete \emph{nodes} arranged 
geometrically in a Bravais lattice \cite{thesis:wylie_1990,book:lattice-gas_hydro_2001,review:dHumieres_Lallemand}. Since the number of nodes is finite, 
suitable boundary conditions (most commonly, periodic) must be implemented.
Each node has a fixed number of \emph{channels} that can be either empty or occupied
by a single \emph{particle}. The channels point to the nearest neighbouring nodes, 
therefore a particle in each channel is considered to possess a single speed in the direction of the neighbouring node to which that channel is pointing.

The time evolution proceeds in discrete steps where each single step consists of 
two phases -- \emph{propagation} and \emph{collision}. During the propagation phase, 
the particles move, i.e. the occupied state of a channel in each node is transferred to 
the channel of the same direction in the corresponding neighbouring node. In the 
collision phase, the states of each node change locally, according to a set of fixed 
rules. 
In order to reproduce the macroscopic properties of a physical fluids, the collision rules and lattice geometry are constructed in such a way that the relevant conservation laws and certain symmetries remain in tact.

The rigid obstacles and impermeable boundaries are introduced by setting up special collision rules describing particle reflection for the nodes at the boundaries. Also 
the sources and sinks may be added by special rules of particle creation/destruction 
at some nodes.

\subsection{The FHP models}
\label{subsection:models_fhp}

The FHP models \cite{article:fhp} belong to a class of two-dimensional 
lattice gas models based on the two-dimensional triangular lattice. There are several versions of the FHP models that maintain the same lattice structure, but differ in collision rules.

The simplest version is the so-called FHP-I model where each node has 6 channels 
corresponding to the 6 directions on the triangular lattice.
For our simulations we have used the FHP-II and FHP-III versions where 7 particles at 
each node may exist -- six moving and one additional particle at rest (having zero velocity). Besides having a higher number of possible effective collisions, the main feature of the FHP-III model compared to FHP-I and FHP-II models is the property of \emph{self-duality}. This means that the dynamics of particles (occupied channels) is equivalent to the dynamics of holes (unoccupied channels) and the dynamics, i.e., 
the collision rules for the dual states (with occupied and unoccupied channels exchanged) are the same as for the original states. 

\subsection{Averaging and macroscopic observables}
\label{subsection:averaging}
The discrete dynamics of the states of nodes on the Bravais lattice described above constitutes the \emph{microscopic} dynamics of the model with microscopic quantities, such as local density (number of particles at the node), velocities of particles (or local velocity field taken as an average velocity of all particles at the given node). These quantities have little to do with actual microscopic dynamics that takes place in real physical fluids.
However, under appropriate circumstances, the \emph{macroscopic} properties of the lattice gas can reproduce the macroscopic properties of real fluids.

The macroscopic observables from the lattice gas simulations are obtained by spatial and temporal averaging. Spatial averaging consists in averaging the microscopic states over blocks of nodes. Temporal averaging means that the value of the state is taken averaged over multiple time steps.

Here we have used spatial averaging over $16 \times 16$ blocks of nodes and for the 
velocity field. Also, temporal averaging over 10 time steps has been used 
{unless noted otherwise}.

\subsection{Reynolds and Strouhal numbers}
\label{subsection:models_reynolds-strouhal}
Reynolds number is a dimensionless number that characterizes the flow by
showing the relative importance of inertial and viscous forces \cite{book:landau_hydro_1987}. It is widely used 
to quantitatively describe different regimes of the flow.

For the FHP-III model, the Reynolds number $Re$ is calculated in the following way \cite{book:lattice-gas_hydro_2001}. First, the density $\rho$ of the particles on the lattice is measured. Since it is a number from 0 to 7 for each node, it is convenient to use the reduced density $d=\rho/7$. 
Because of the self-duality of the FHP-III model, if $d>0.5$, then the dynamics of holes instead of particles is being observed, therefore  $d\equiv 1 - d$ in that case. Because of this, certain macroscopic observables differ 
 from theoretical ones by the density dependent non-Galilean factor \cite{book:lattice-gas_hydro_2001}:
 
\begin{equation}
	g(d)=\frac{7}{12}\frac{1-2 d}{1 - d}\,\,.
	\label{eq:non-galilean}
\end{equation}
 
Another important quantity required in order to calculate $Re$ is the kinematic viscosity:

\begin{equation}
	\eta = \frac{1}{28 d \bar{d}(1-\frac{8}{7}  d \bar{d})}-\frac{1}{8}
	\label{eq:viscosity}
\end{equation}
where $\bar{d}=1-d$. The Reynolds number is
\begin{equation}
	Re=\frac{g u L}{\eta}\,\,.
	\label{eq:reynolds}
\end{equation}
where $u$ is the average velocity magnitude and $L$ is the typical dimension of the obstacle. One readily notices, that in order to increase $Re$, one has to choose a wide channel, produce high velocity of the flow and optimize $d$.
In the present case $Re$ is maximized at $d\cong 0.305$.

The Strouhal number $St$ is another dimensionless quantity, characterizing the flow. The function $St(Re)$ provides important information about what is happening at the wake \cite{book:lattice-gas_hydro_2001}. It is defined by the following equation:

\begin{equation}
	St = \frac{f\cdot L}{u\cdot g}\,\,,
	\label{eq:strouhal}
\end{equation}
where $L$ is again the typical size of the obstacle (in this case, the length of the base of the triangle, see Fig.~\ref{fig:geo_both}). $f$ is the frequency of the wake oscillation produced by vortex shedding. All the quantities are in natural lattice units (i.e. number of lattice sites and time steps).

The industrial vortex flow meters function under the assumption of constant $St$. If this were the case, the frequency $f$ of the vortex shedding would depend linearly on the flow velocity $u$. However this turns out not to be true at least in some regimes of the flow \cite{article:reik_et_al_2010}.

In our case $f$ is determined by the lattice gas hot-wire anemometry \cite{book:lattice-gas_hydro_2001} -- averaging over a block where the local velocity magnitude is recorded at each step. Later on, Fourier analysis and sine function fitting is used to determine the low frequency mode from the noisy signal, since a direct application of the FFT is often not a good option, as it requires many steps to obtain reasonably small errors
{due to fluctuations of the velocity field, which in turn
come about due to complex flow patterns}. The low frequency mode in the case of K\'arm\'an vortex street corresponds to the vortex shedding frequency. A piece of raw data and the corresponding sine function fit are provided in Fig.~\ref{fig:fourier}.

\section{Simulation of a vortex flow meter}
\label{section:simulation}

{
Even though there is a considerable body of knowledge concerning vortex formation in flows behind
cylinders \cite{roshko}, and other highly symmetric objects in translationally-invariant
geometries, the particular case of a prism in a confined setting has not been studied yet 
up to the best of our knowledge.}
Since the main goal of this paper is to simulate the vortex flow meter that is usually placed in a pipe, as in \cite{article:reik_et_al_2010}, periodic boundary conditions have been used in the $x$ direction being main direction of the flow (from the left to to the right in the figures), and the containment of the flow by the pipe walls has been implemented as impermeable boundaries from the top and the bottom (i.e., in $y$ direction).

This section presents results from a series of simulations in several different
geometries. First, the velocity profile of the steady flow without an obstacle has been obtained in order to test the velocity profile (Fig.~\ref{fig:fan_source}). Then, the vortex shedding from a triangle has been implemented and visualized in both FHP-II and FHP-III models (Fig.~\ref{fig:fhp2_3}).
Finally, the two-dimensional model of a vortex flow meter has been simulated by placing 
the blunt prism-shaped vortex shedding device (Fig.~\ref{fig:geo_both}) in the flow with two different ratios of obstacle size to channel width in order to measure the dependence of the vortex shedding frequency on the flow velocity and determine the Strouhal-Reynolds number dependence (Fig.~\ref{fig:results_geo1} and Fig.~\ref{fig:results_geo2}).

\subsection{Velocity profile of the laminar flow}
\label{subsection:simulation_laminar}
Before starting the simulation of unsteady flow of the vortex street 
behind an obstacle, the velocity profile of an unobstructed flow has been 
investigated using the FHP-II model. 
On a lattice of $120\times 48$ cells (each cell, as mentioned before, being a block of  $16 \times 16$ nodes) the velocity component $v_x$ along the general flow direction has been measured. 
A steady-state velocity profile has been determined for every horizontal block of cells  (coordinate $y$ ranging from 1 to 48) by spatial averaging of the velocity over cells 40 to 100 in the $x$ direction and temporal averaging over 100 time steps.
Two mechanisms of flow induction have been considered. 

\begin{figure}
  {\centering
  \includegraphics[width=0.45\textwidth]{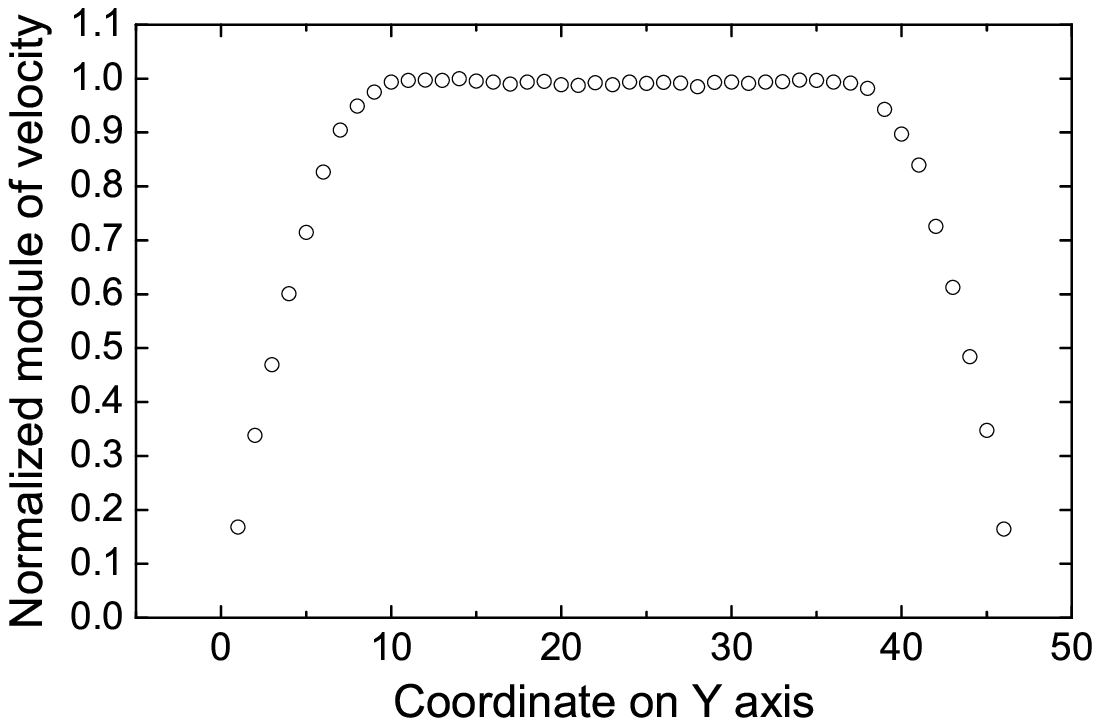}
  \includegraphics[width=0.45\textwidth]{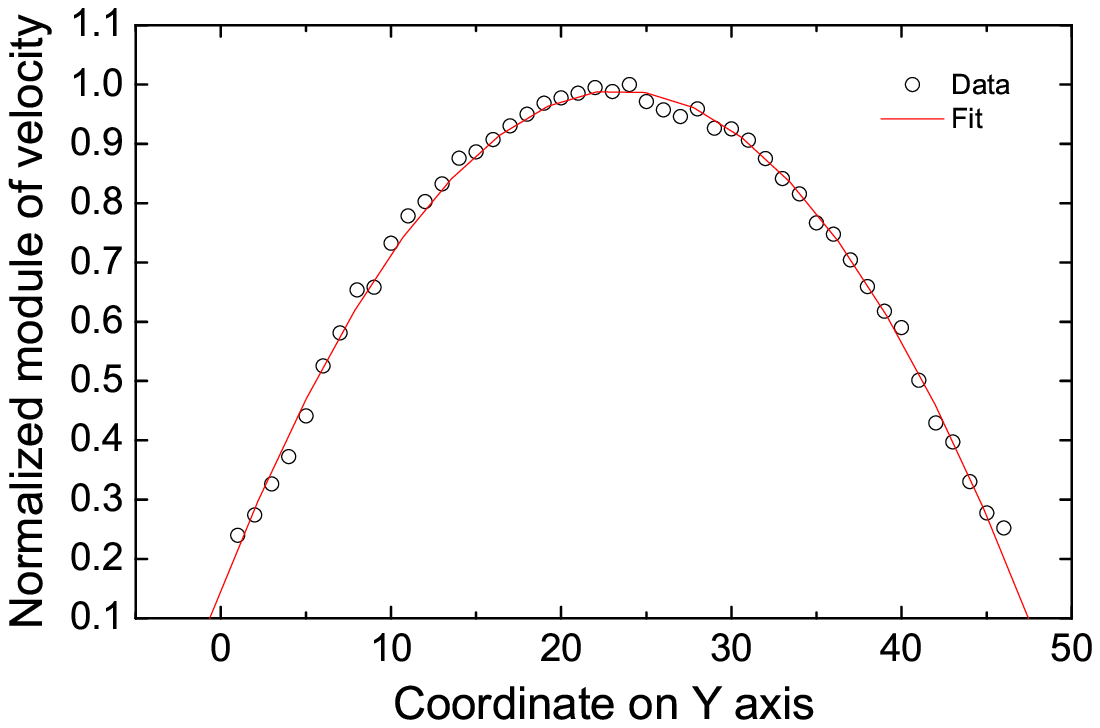}\par}
  \caption{Left panel: rectangular velocity profile resulting from the fan approach. Right panel: velocity profile in the source/sink case. Circles represent the measured data, and the solid red line is the Poiseuille profile fit.}
  \label{fig:fan_source}
\end{figure}

First, the so-called fan approach \cite{thesis:wylie_1990} has been implemented. This approach consists of 
a vertical zone of $1 \times 48$ cells where each particle moving to the left (in the opposite direction to $x$) is being reversed with probability $0.001$. Using this approach, however, an almost rectangular velocity profile has been observed (left panel of Fig.~\ref{fig:fan_source}), instead of the expected Poiseuille profile \cite{book:monin1}.

Next, we have used the source/sink flow induction mechanism. A source or a sink is 
a node where each arriving particle is absorbed (destroyed) and new particles moving in all the available directions (6 in the FHP case) are introduced each with some probability \cite{thesis:wylie_1990}. If, for example, this probability is $0.2$, then $6\times 0.2 =1.2$ particles at the source/sink node are created on average. If, on average, there are more particles produced than destroyed, then such a node acts as a source, and, if there are more particles destroyed than created, a node acts as a sink.

We have implemented the source/sink flow induction by introducing two vertical zones of $1\times 48$ of source/sink cells at the opposite sides of our system with different particle creation probabilities ($0.5$ and $0.4$ in this case). After a longer equilibration period of about 20000 time steps, the expected Poiseuille velocity profile has been observed (see right panel of Fig.~\ref{fig:fan_source}). 
Therefore, the source/sink induction of the flow has been used for further simulations.

\subsection{Vortex shedding in FHP-II and FHP-III versions}
\label{subsection:simulation_fhp2vs3}

\begin{figure}
  {\centering
  \includegraphics[width=0.45\textwidth]{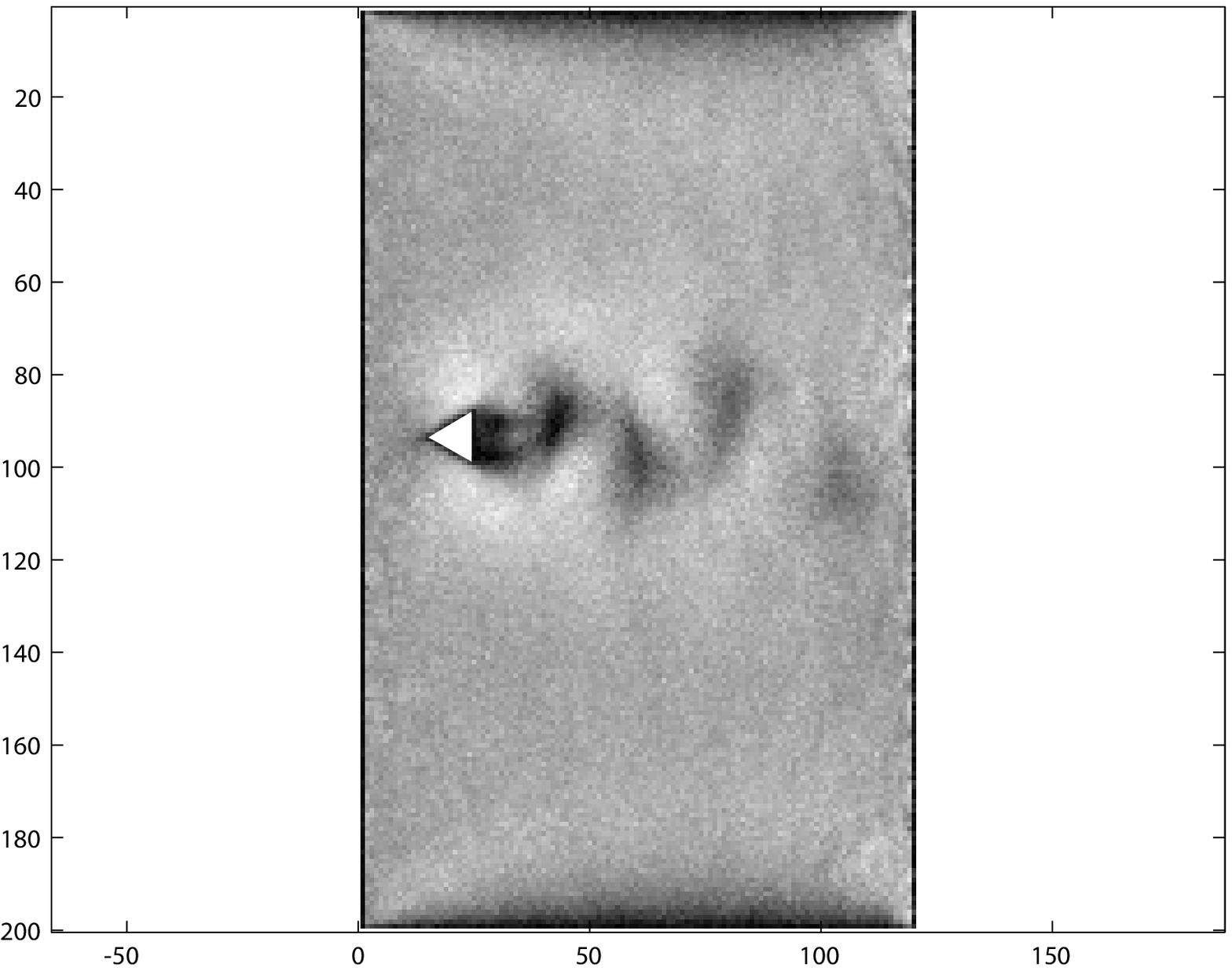}
  \includegraphics[width=0.45\textwidth]{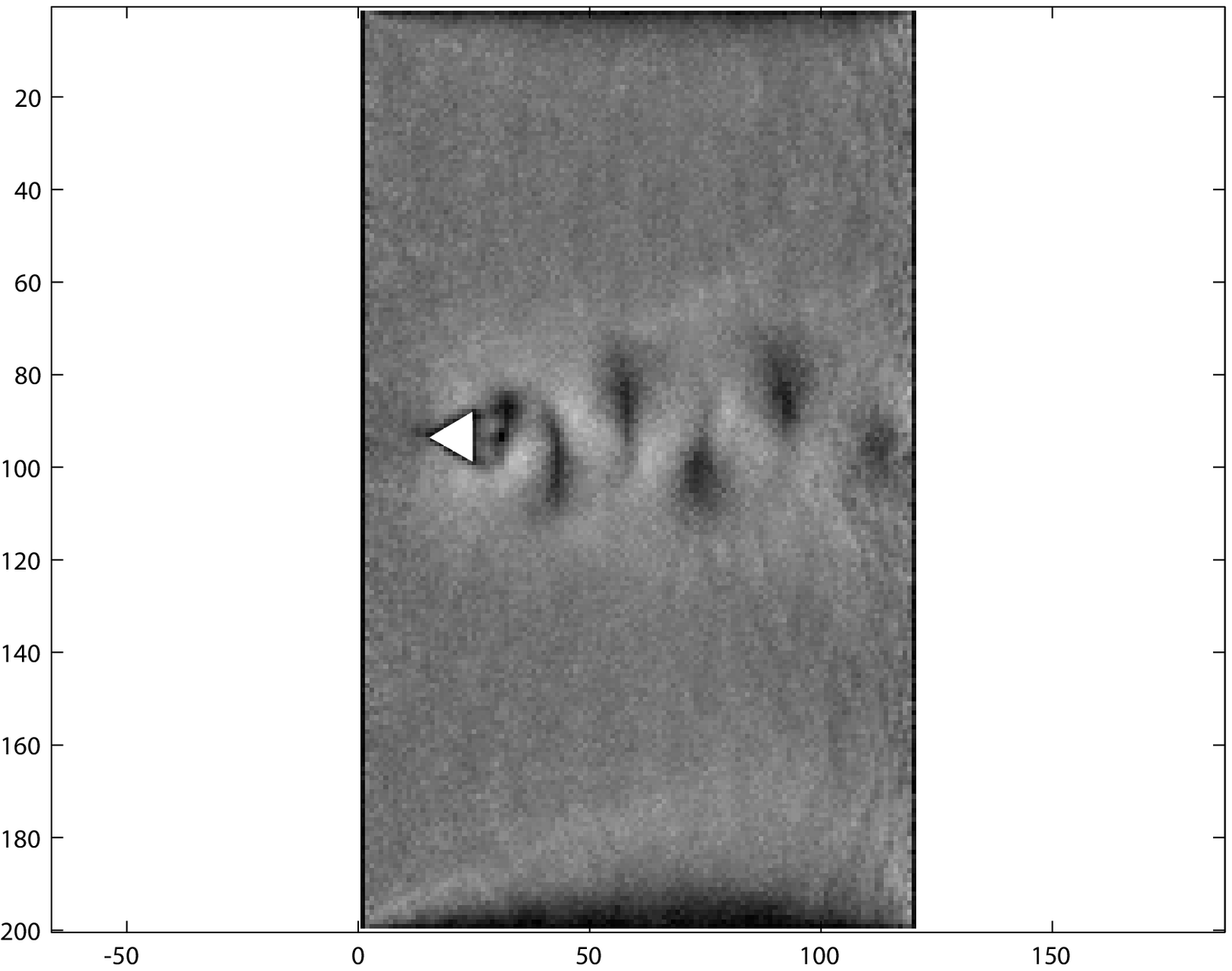}\par}
  \caption{Wake of a triangular obstacle at source-sink ratio 0.6/0.1 for two versions of FHP. The velocity magnitude is represented by different shades of gray. White bitmap image of the triangular object has been placed by hand on top of the calculated velocity field. Left panel: FHP-II. Right panel: FHP-III.}
  \label{fig:fhp2_3}
\end{figure}

The numerical scheme has been tested further by comparing the vortex shedding in the FHP-II and FHP-III models. For this, we have introduced a solid obstacle shaped as an equilateral triangle in the flow (Fig.~\ref{fig:fhp2_3}).

The creation probabilities of $0.6$ and $0.1$ of the source/sink zones have been used, and a simulation of 30000 steps has been carried out on a lattice of
120x200 blocks. 
The absolute magnitude of the velocity
displayed in 100 shades of gray (white being highest magnitude) is shown. Note that the K\'arm\'an vortex street is clearly visible in both FHP-II and FHP-III models 
(left and right panels of Fig.~\ref{fig:fhp2_3}, respectively).
However, one can also notice that FHP-III produces more pronounced vortices than FHP-II, owing to the lower viscosity and therefore a higher Reynolds number \cite{review:dHumieres_Lallemand}. 
The lower viscosity of the FHP-III model stems from
its expanded set of possible collisions. In the
latter, 76 configurations participate in collisions,
as opposed to merely 22 such active configurations
of the FHP-II model. We refer the reader to
Refs.~\cite{review:dHumieres_Lallemand} and \cite{thesis:wylie_1990}, where these collisions
are listed explicitly.
For this reason, we consider the FHP-III rule set to be more suitable for the measurement of the Strouhal number, as it produces clearer vortices with no additional computational effort. Thus, the FHP-III model has been used for the simulations of the vortex flow meter.

\subsection{Vortex flow meter}
\label{subsection:simulation_flowmeter}

The main part of our investigation consists of measurements of the vortex shedding frequency dependence on the flow rate in the two-dimensional simulation of a vortex flow meter \cite{article:reik_et_al_2010} using the FHP-III rule set. The results have then been used to determine the Reynolds-Strouhal number dependence.

\begin{figure}
	{\centering\includegraphics[width=0.4\textwidth]{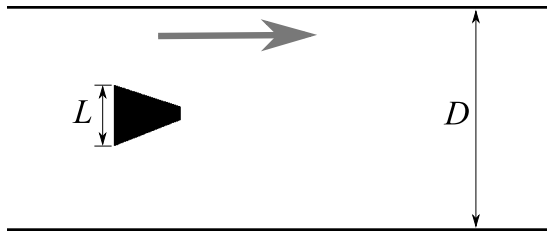}\par}
  \caption{Geometry of the vortex flow meter simulation.}
	\label{fig:geo_both}
\end{figure}

The two-dimensional model of a vortex flow meter consists of a flow in a channel with impermeable walls and a trapezoid-shaped obstacle that constitutes the vortex shedding device.
We have considered two cases differing in the obstacle to channel size ratio, i.e., the ratio between the length $L$ of the longer base of the vortex shedding device and the width $D$ of the channel. All simulations used a geometrically similar vortex shedding device with the length of the shorter base and the height of the trapezoid proportional to $L$ and equal to $0.225\,L$ and $1.1\,L$, respectively. The geometry is depicted in Fig.~\ref{fig:geo_both}. Here, the general flow direction is indicated by the gray arrow. 

\begin{figure}
	{\centering\includegraphics[width=0.4\textwidth]{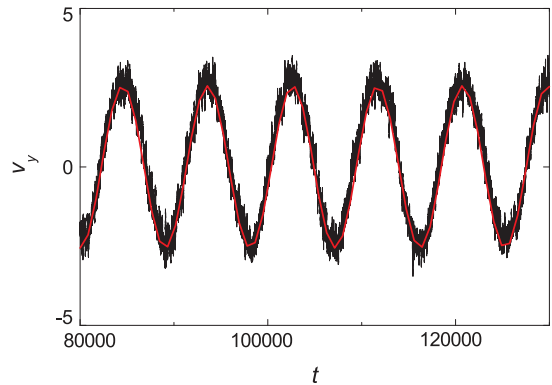}\par}
  \caption{Dynamics (in time steps) of the velocity perpendicular to the channel direction used for determination of the wake oscillation frequency. Here the case of the small vortex shedding device at small flow velocity $u=0.267$ ($Re=156$) is shown. The black line represents the output and the red line represents the sine fit for the frequency $0.0001$.}
	\label{fig:fourier}
\end{figure}

 We have investigated a relatively small vortex shedding device with obstacle to channel ratio $L/D=0.055$ and a large vortex shedding device with $L/D=0.27$. 
{
This particular choice of the two ratios has been made 
for two reasons. First, these choices address the
two opposite physical limits: (i) the 
transparent situation
where the vortices are shed far from the walls
of the pipe
($L/D=0.055\ll 1)$, and 
also (ii) the less intuitive case where the boundary effects
should play an important role ($L/D=0.27\sim 1$).
Moreover, the number $0.27$ has been read off the geometry
of the industrial vortex flow meter 
investigated in Ref.~\cite{article:reik_et_al_2010}, in order
to make a contact with the results presented there.
}
 In both cases the $y$ component of the velocity $v_{y}$ (velocity in the direction perpendicular to the channel flow direction) has been measured 5 cells downstream from the shorter base of the trapezoid that constitutes our vortex shedding device.
This corresponds to the lattice-gas implementation of 
hot-wire anemometry. 
Vortex shedding has produced sine-shaped variation in $v_{y}$.
 An example piece of raw data that has been measured in the simulation is depicted in Fig.~\ref{fig:fourier} together with the sine function fit from which the vortex shedding frequency is determined.

The Reynolds number $Re$ and the Strouhal number $St$ have been calculated from the measured flow velocity $u$ and the vortex shedding frequency $f$ using (\ref{eq:reynolds}) and (\ref{eq:strouhal}), respectively. The velocity $u$ has been tuned by changing the source/sink ratio of the particle-absorbing/producing zones described in Subsection $\ref{subsection:simulation_laminar}$. It has been measured by averaging across the channel upstream from the obstacle.

\subsubsection{Small vortex shedding device}
\label{subsubsection:simulation_flowmeter_small}

\begin{figure}
  {\centering
  \includegraphics[width=0.47\textwidth]{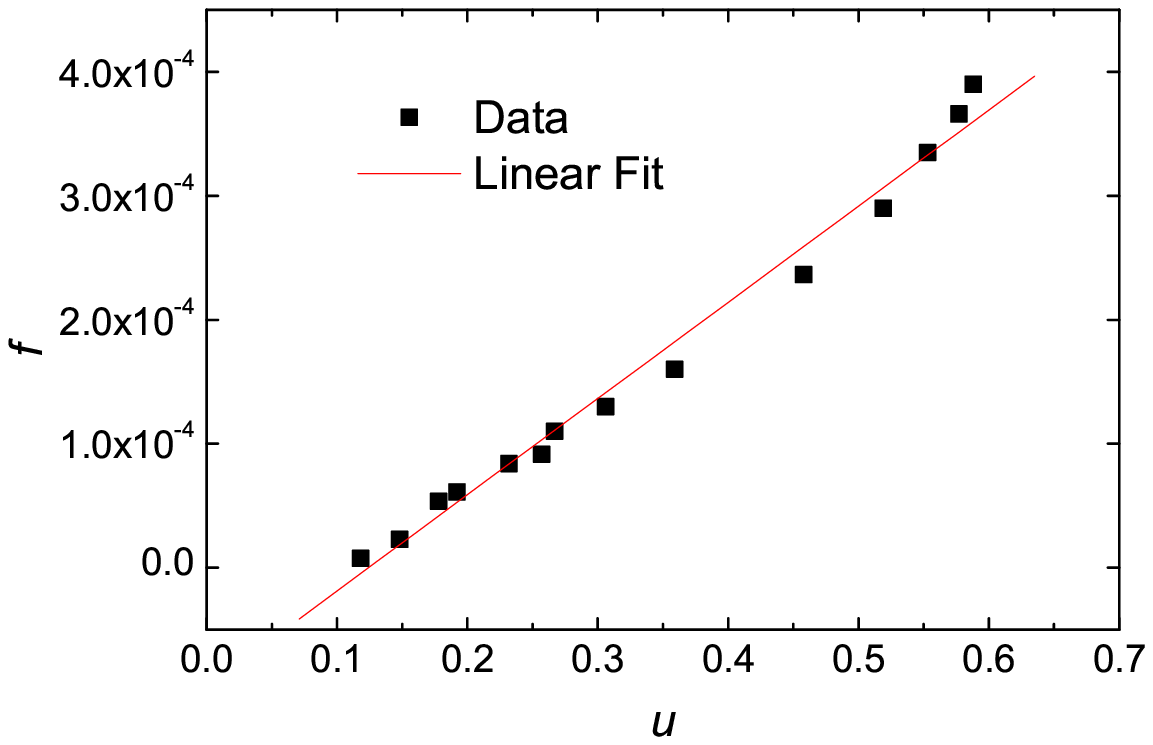}
  \includegraphics[width=0.43\textwidth]{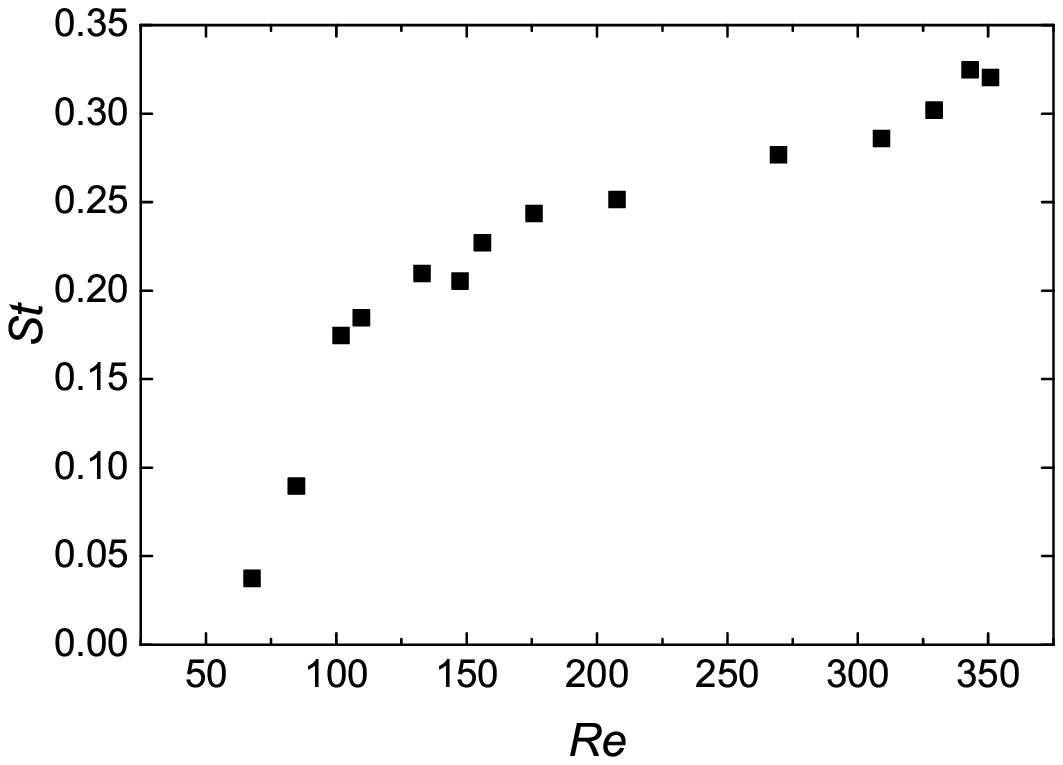}\par}
  \caption{Measurements for the small vortex shedding device ($L/D=0.055$, see Fig.~\ref{fig:geo_both}). Left panel: vortex shedding frequency dependence on flow velocity with linear fit (solid red line). Right panel: Strouhal-Reynolds number dependence.}
  \label{fig:results_geo1}
\end{figure}

For the simulation with a small vortex shedding device where $L/D=0.055$, a system of $120\times 200$ cells (each cell, as before, being a $16\times 16$ block of lattice nodes) has been used. The measured vortex shedding frequency $f$ dependence on the flow velocity $u$ and the computed $St(Re)$ dependence are shown in Fig.~\ref{fig:results_geo1}.

An approximate linear dependence of $f$ on the velocity $u$ has been observed:
\begin{equation}
	f=-(1\pm 0.1) \cdot 10^{-4} + (7.8 \pm 0.3) \cdot 10^{-4} \cdot u
	\label{eq:lin_f}
\end{equation}
The uncertainties here are the errors in the least-squares linear fit of the data.
 
However, one notices that the linear dependence is not ideal. First of all, it would give a non-zero frequency $f$ for $u=0$. Secondly, one can see a nonlinear trend in the data (see left panel of Fig.~\ref{fig:results_geo1}) which suggests that $St$ is not constant. The latter fact is clearly visible when looking at the $St(Re)$ dependence computed from the data (see right panel of Fig.~\ref{fig:results_geo1}).

\subsubsection{Large vortex shedding device}
\label{subsubsection:simulation_flowmeter_large}

The authors of Ref.~\cite{article:reik_et_al_2010} have observed the decrease
in Strouhal number $St$ with increasing Reynolds number $Re$ for $Re<4800$ and
suggested that the reason for this dependence might be related to the formation
of horseshoe vortices along the channel walls and the three-dimensional
turbulent flow. 

{
Both of these effects are specific for a three-dimensional geometry,
and therefore do not exist in our flat model. It is thus
useful to study if the previously reported trend in the $St(Re)$ dependence
survives given the decreased dimensionality of the system.
}

\begin{figure}
  {\centering
  \includegraphics[width=0.46\textwidth]{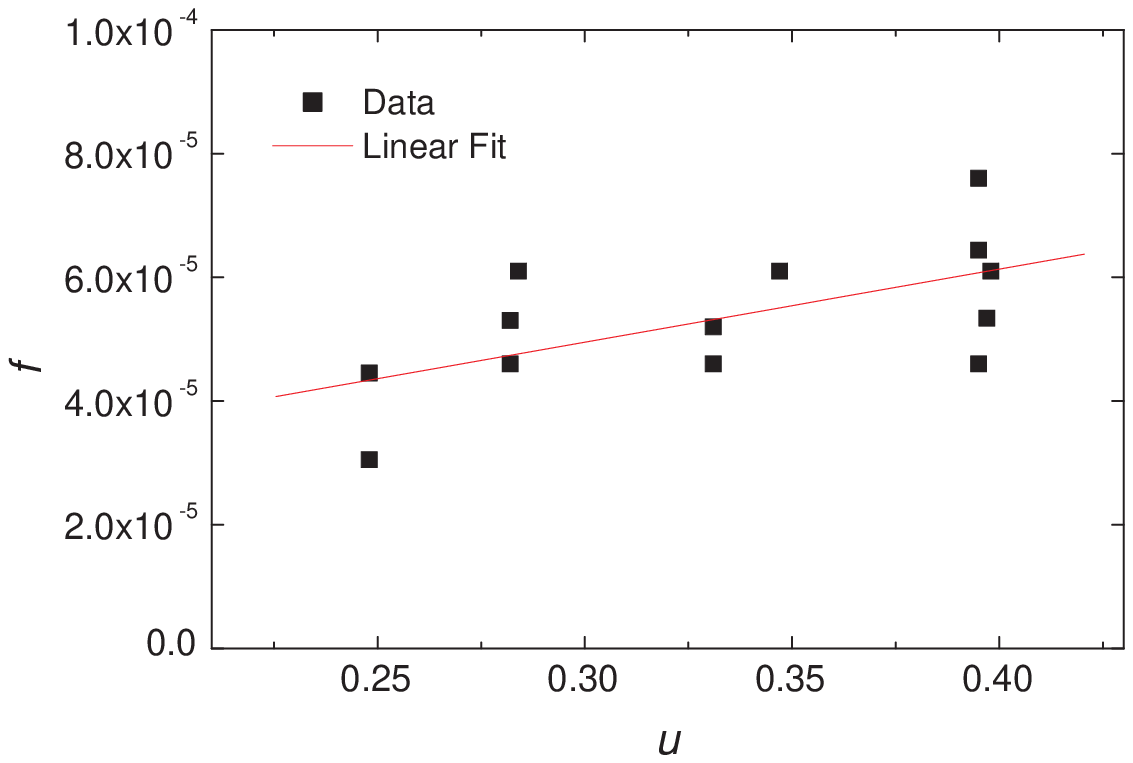}
  \includegraphics[width=0.44\textwidth]{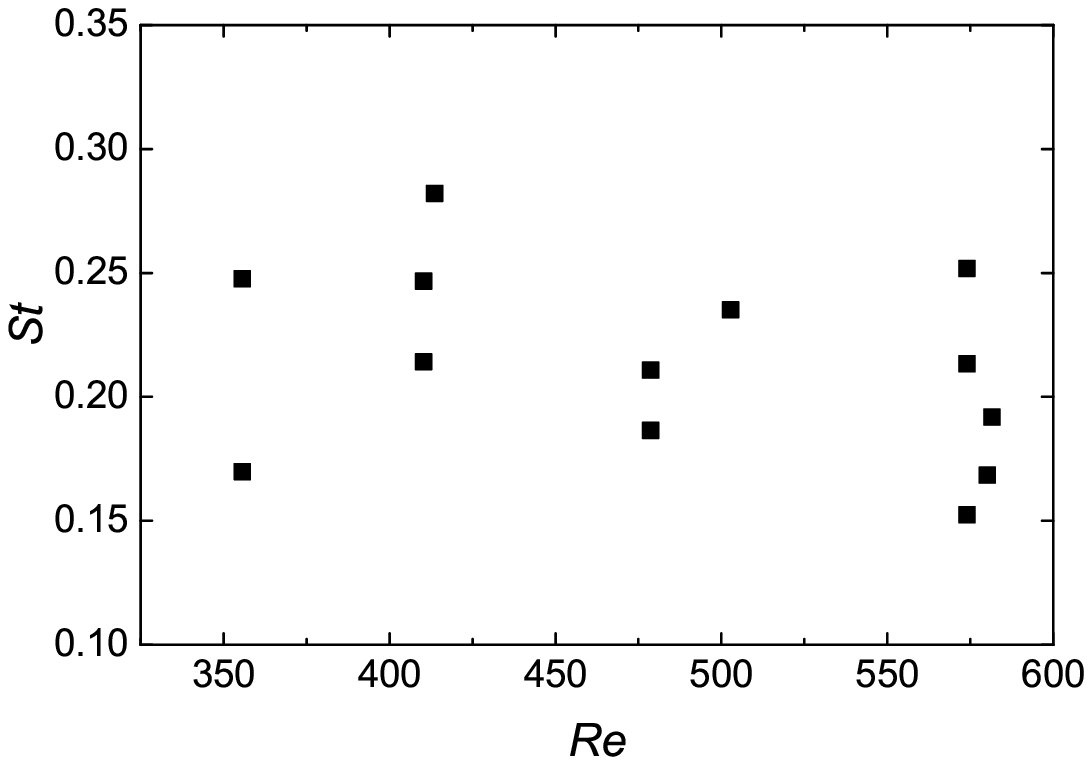}\par}
  \caption{Measurements for the large vortex shedding device ($L/D=0.27$, see Fig.~\ref{fig:geo_both}). Left panel: vortex shedding frequency dependence on flow velocity with linear fit (solid red line). Right panel: Strouhal-Reynolds number dependence.}
  \label{fig:results_geo2}
\end{figure}

Results for the simulation of a large vortex shedding device with $L/D=0.27$ are shown in Fig.~\ref{fig:results_geo2}. Here, a system of $240\times 100$ cells has been used. The ratio $L/D=0.27$ has been chosen to be the same as has been used in Ref.~\cite{article:reik_et_al_2010} thus allowing a direct comparison of our results with the ones presented in Ref.~\cite{article:reik_et_al_2010} for the three-dimensional case. Due to the interaction between the vortex street and the channel walls, a very noisy signal has been obtained. The dependence of the frequency $f$ on $u$ might still be considered as slightly increasing (left panel of Fig.~\ref{fig:results_geo2}):
\begin{equation}
	f=-(0.1\pm 0.1) \cdot 10^{-4} + (1.2 \pm 0.5) \cdot 10^{-4} \cdot u\,\,,
	\label{eq:lin_f2}
\end{equation}
but no clear trend in the dependence of $St$ on $Re$ is apparent (right panel of Fig.~\ref{fig:results_geo2}).

\section{Summary and conclusions}
\label{section:conclusions}
In summary, this paper presents three main results from a series of two-dimensional hydrodynamic simulations using the FHP lattice gas models. 

First, the Poiseuille profile for the laminar flow confined in a channel has been demonstrated using the source/sink method. It has also been shown that the so-called fan approach for induction of the flow results in a different, namely, rectangular, velocity profile (Fig.~\ref{fig:fan_source}).

Moreover, the vortex shedding from a triangular object in the flow has been demonstrated in FHP-II and FHP-III models, exhibiting higher viscosity in the FHP-II model (Fig.~\ref{fig:fhp2_3}). Therefore, the FHP-III model has been used for further simulations.

The main part of this paper has been the simulation of the vortex shedding from a blunt trapezoid-shaped obstacle (Fig.~\ref{fig:geo_both}) in a confined flow. This configuration
is a model for 
the vortex flow meter described in Ref.~\cite{article:reik_et_al_2010}. The Strouhal-Reynolds number dependence was investigated in two different obstacle-channel size ratios.

{
As already noticed in classical works (see, e.g.,~Ref.~\cite{article:ssb88prl}), statistical fluctuations play
a prominent role in lattice gas automaton simulations
in general, and in turbulence-related problems in particular.
Having this limitation of our method in mind, we have
performed the simulation multiple times in order
to investigate run-to-run noise. We have discovered
that the differences between
runs are appreciable only for the large vortex shedding
device case. Hence, 
we only show the results of different runs for that
case (see Fig.~\ref{fig:results_geo2}). However, in order
to fully ascertain that the results are not dependent
on the statistical fluctuations, one should turn to
more sophisticated methods (see Refs.~\cite{article:mz88prl,article:hsb89epl}).
}

Linear dependence (see Eq.~(\ref{eq:lin_f})) of the vortex shedding frequency on the flow velocity and increasing Strouhal number with increasing Reynolds number has been demonstrated (Fig.~\ref{fig:results_geo1}) for the small vortex shedding device. 

For the large vortex shedding device, where the vortex street is obstructed by the channel walls, only a weak dependence of the vortex shedding frequency on the flow velocity can be observed (see Eq.~(\ref{eq:lin_f2}), Fig.~\ref{fig:results_geo2}) and no significant Strouhal-Reynolds dependence has been found in contrast to the experimental data and hydrodynamic simulations given in Ref.~\cite{article:reik_et_al_2010}.
{ Therefore, our two-dimensional results support 
the hypothesis presented in Ref.~\cite{article:reik_et_al_2010}, namely,
that flow structures particular to the three-dimensional geometry
are responsible for the strong $St(Re)$ dependence.}

{
In future work, it would be interesting to study the
transition from two dimensions to three dimensions, as the onset of the
strong
$St(Re)$ dependence is expected to occur when the extent of the
smallest dimension of the system surpasses the length scale characteristic to the
flow. Therefore, in pipes smaller than the size of the horseshoe vortex
(given a certain flow velocity),
vortex flow meters operate in the accurate linear regime, whereas when the
diameter of the pipe is sufficiently large, the accuracy of the
said flow meters should decrease. These investigations might lead to
a better understanding of the reliable-operation bounds of the
industrial vortex flow meters.
}

\section{Acknowledgements}
\label{section:acknowledgements}

{
It is our pleasure to thank G.~T.~Barkema for introducing us to the
FHP models. We also thank J.~Bu\v{c}inskas for his spirited encouragement
to publish our results. J.~A. was supported by European Union's Horizon 2020 research and innovation
programme under the Marie Sk\l odowska-Curie grant agreement No 706839 (SPINSOCS).
}

\end{document}